\documentclass[12pt]{article}
\usepackage{amssymb,amsmath,cite,bm}
\begin{document}

\title{\bf Energy Efficient Control of Electric Motors}

\author{Farhad Aghili\thanks{email:faghili@encs.concordia.ca}}

\date{}
\maketitle

\begin{abstract}
This paper presents development of  an optimal feedback linearization control (OFLC) for interior permanent magnet (PM) synchronous machines operating in a non steady-sate operating point, i.e., varying torque and speed, to achieve precision tracking performance and energy saving by minimizing the copper loss. An isomorphism mapping between  the dq-axes phase voltages and two auxiliary control inputs over full ranges of torque and speed is established by the linearization controller using the notion of orthogonal projection. The auxiliary control inputs are defined to be exclusively responsible for torque generation and power consumption. Subsequently, an analytical solution for the optimal-linearization control is derived in a closed-form by applying the Hamiltonian of optimal control theory in conjunction with the Pontryagin's minimum principle.  The optimal controller takes the maximum voltage limit and torque tracking constraint into account while maximizing machine efficiency for non-constant operational load torque and speed. Unlike the convectional quadratic regulator-based control of electric motors, the proposed control approach does not rely on steady-state operation conditions and hence it is suitable for such applications as electric vehicles and robotics. 
\end{abstract}

\section{Introduction and motivation}
The market for motor drives  encompasses a wide range of applications areas from consumer appliances and electric cars to industrial robotics and machine tools. Precise and fast torque tracking performances and energy-efficiency over the entire speed/torque range of motors is highly required in many of these applications \cite{Straete-Degezelle-DeSchutter-1998,Aghili-Buehler-Hollerbach-2001}. PM synchronous motors (PMSMs) are often suitable choices for these applications as they exhibit high torque-to-mass ratio, fast dynamic responses, and high power efficiency compared to other types of motors. In fact, the recent surge in demand for high efficient PM motor drives is mainly caused by rapid proliferation of motor drives into the automobile industry.
The performance and efficiency of  PMSMs are constantly being improved through not only optimal motor design and  construction but also by implementation of advanced control methods \cite{Grenier-Dessaint-Akhrif-1997,Zhong-Rahman-Hu-1997,Aghili-Buehler-Hollerbach-2007,Buja-Kazmierkowski-2004,Aghili-2015a,Aghili-2017b}  for electric vehicles as well as industrial automation and robotics within the manufacturing sector.  Such applications often require motor drives to work in wide speed and torque ranges  while maintaining high efficiency \cite{Cheng-Tesch-2010}.

The is a variety of control methods in literature for computer control of electric machines. Minimization of power dissipation has been considered in some of these control approaches   for electric machines working in steady-state operation conditions, i.e., constant torque or speed. A modification to the conventional vector-controlled drive system for special Synchronous machines is  proposed in \cite{Sue-Pan-2008} to minimize copper losses based on a voltage-constraint-tracking in the field-weakening control. However, the control is designed based on steady-state voltage equations of a machine in the rotor reference frame, i.e., time-derivatives of currents are ignored. There are also other direct control possibilities such as  feedback linearization control (FLC) \cite{Bodson-Chiasson-Novotnak-1993,Kaddouri-Akhrif-1994,Grenier-Dessaint-Akhrif-1997} or direct torque control (DTC) \cite{Takahashi-Noguchi-1986,Zhong-Rahman-Hu-1997,Zhang-Zhu-2011,Gulez-Adam-Pastaci-2007,Cho-Yongsoo-Kim-2013}. The DTC schemes  have been further developed  to minimize copper loss or to defer voltage saturation using flux-weakening control in order to extend the range of operational speed of PMSMs \cite{Chen-Chin-2000,Zhu-Chen-How-2000,Chen-Chin-2003,Chiang-Cheng-Hsum-2015}.
A torque regulator and a flux regulator  for PMSMs operating in a wide speed range requiring high torque/power accuracy and a fast
dynamic response is proposed in \cite{Cheng-Tesch-2010}. Linear quadratic regulator (LQR)-based optimal vector control of PMSM in dq-axes reference frame using linear state feedback is reported in \cite{Chang-Low-1994,Do-Kwak-2014,El-Sousy-2013,Cheema-Fletcher-2016}. Again, these optimal control approaches are designed for troque/speed regulation of PMSMs working in steady-state operating point.
It is worth noting that pulse width-modulation inverter
nonlinearities affect high-frequency carrier-signal voltage and that motivated researchers to developed a  number of techniques to compensate for inverter's nonlinearities \cite{Guerrero-Leetmaa-Briz-2003,Kazmierkowski-Malesani-1998,Sepe-Lang-1994} or limited bandwidth \cite{Aghili-2011m}. It is also known that the motor parameters are also prone to change in time mainly due to temperature changes and that can cause performance-degradation of the control system. On-line identification schemes have been adopted for control of PMSMs to cope with the parametric uncertainty~\cite{Krneta-Antic-Stojanovic-2005,Aghili-2008a,Obeidat-Wang-Lin-2013}. Alternatively, sliding mode observer or sliding model control techniques can be adopted for control of PMSMs to achieve fast response and robustness w.r.t. the effect of the variations of motor parameters \cite{Li-Elbuluk-2001,Reichhartinger-2010,Qiao-Shi-Wang-2013}.

Optimal-current determination
for multiphase PMSMs in real time are reported in
\cite{Jovanovic-Betz-1999,Aghili-Buehler-Hollerbach-2000b,Mademlis-Xypteras-2000,Aghili-Buehler-Hollerbach-2003,Aghili-2011m,Kestelyn-Semail-2011}.
These indirect optimal torque
controller typically requires current source inverter equipped with a large bandwidth current controller in order to be able to inject currents into the inductive windings without introducing significant phase lag and
for smooth torque production. The armature current vector can be controlled using Maximum Torque Per Ampere (MTPA) control technique  for maximizing torque for a given magnitude of current vector  \cite{Jahns-Kliman-1986,Nakai-Ohtani-Satoh-2001,Morimoto-Hatanaka-Tong-1991}. High-performance current regulator to improve the current responses  in  high-speed flux-weakening region  by a feedforward compensator is developed in \cite{Morimoto-Sanada-Takeda-1994}. This control strategy has been widely adopted in constant torque operating range  to achieve fast transient and high-efficiency operation of PMSM drive systems \cite{Nakai-Ohtani-2005,Sue-Pan-2008}.
A control scheme for a PMSM drive that can automatically switch to the MTRA or the field-weakening control modes
is proposed in \cite{Sue-Pan-2008} to maintain minimum copper-loss operation. The switching controller relies on an empirical method to properly select the values of a couple of tuning parameters for a stable operation   \cite{Sue-Pan-2008}. MTPA control scheme for PM motors considering the temperature dependence of the magnet field and saturation effect is proposed in~\cite{Rang-Lim-Nam-2004}. These MPTA controllers require current source power converters  which amplify the  instantaneous phase current commands by means of phase current feedback to provide high-bandwidth closed-loop current regulation \cite{Jahns-Kliman-1986}.

This work applies the Hamiltonian of optimal control theory to achieve optimal feedback linearization control (OFLC) of PM synchronous machines operating with varying speed/torque \cite{Aghili-2017b}. A feedback linearization scheme with two distinct control inputs is developed based on notion of orthogonal projection. The stationary dq-axes phase voltages are related to two auxiliary control inputs through an isomorphism mapping in such a way that the control inputs are exclusively responsible for torque generation and energy minimization. Subsequently, an optimal controller is developed based on the {\em maximum principle} formulation for minimize power consumption subject to accurate torque production and voltage limits.  The important features of the optimal feedback linearization  scheme is that it works for time-varying torque or
variable-speed drive applications.

\section{Generalized linearization model of synchronous Motors} \label{sec:SpatialFreqMotor}
The voltage equations of synchronous motors with salient-pole can be written in the $d$, $q$ reference frame by
\begin{subequations} \label{eq:vdq}
\begin{align} \label{eq:vLdiffi}
L_d \frac{d i_d}{d t} & = - R i_d + L_q i_q \omega + v_d\\
L_q \frac{d i_q}{d t} & = - R i_q + L_d i_d \omega - \psi \omega + v_q
\end{align}
\end{subequations}
where $L_q$ and $L_d$ are the $q-$ and $d-$axis inductances, $i_q$, $i_d$, $v_q$, and $v_d$ are the $q-$ and $d-$axis currents and voltages, respectively, $\psi$ is the motor back EMF constant, and $\omega$ is motor speed \cite{Sebastian-Slemon-Rahman-1987,Pillay-Krishnan-1989,Gebregergis-Chowdhury-2015}. The equation of motor torque, $\tau$, can be described by
\begin{equation} \label{eq:tau}
\tau = \frac{3}{2} p \big(\psi i_q +(L_d -L_q) i_d i_q \big),
\end{equation}
where $p$ is the number of pole pairs \cite{Pillay-Krishnan-1989}. Suppose
\begin{equation}
\eta = \frac{L_q}{L_d} -1 \quad \mbox{and} \quad \mu = \frac{L_q}{R}
\end{equation}
are defined as the inductance ratio and  the machine time-constant, respectively, and vector $\bm i =[i_d \;\; i_q]^T$ contains the stationary dq-axes phase currents. Then, the time-derivative of \eqref{eq:tau} takes the form
\begin{equation} \label{eq:dot_tau}
\dot{\tau} = \frac{3}{2} p \big( \psi \dot i_q +(L_d -L_q) (\dot i_d i_q + i_d \dot i_q)  \big),
\end{equation}
After substituting the time-derivative of the dq-axes phase currents from \eqref{eq:vdq} into  \eqref{eq:dot_tau} and rearranging the resulting equation, we arrive at
\begin{equation} \label{eq:dottau}
\tau + \mu \dot{\tau} = \bm b^T(\bm i)  \bm v + \phi(\bm i,\omega)
\end{equation}
where
\begin{align*}
\bm b(\bm i) = &[ b_d \;\; b_q]^T,\\
b_d(\bm i) = & -\frac{3p}{2R} \eta {L_d} i_q, \\
b_q(\bm i) =&  \frac{3p}{2R} (\psi- \eta L_d i_d),\\
\phi(\bm i, \omega) = & \frac{3p}{2} \frac{\omega}{R} \big( L_q i_d \psi - \eta L_q^2 i_q^2 - \eta  L_d^2 i_d^2 - \psi^2  \big) + \frac{3p}{2} \eta L_q i_d i_q
\end{align*}
Notice that the motor phase currents $i_a, i_b$, and $i_c$ are related to stationary dq-axes phase currents by
\begin{equation} \label{eq:Park}
\begin{bmatrix} i_d \\ i_q \end{bmatrix} = \bm K(\theta) \begin{bmatrix} i_a \\ i_b \\ i_c \end{bmatrix}
\end{equation}
where $\theta$ is the mechanical angle and 
\begin{equation} \notag
\bm K(\theta) = \frac{2}{3}\begin{bmatrix} \cos(p \theta) &  \cos(p \theta - \frac{2\pi}{3}) & \cos(p \theta +\frac{2\pi}{3}) \\ \sin(p \theta) &  \sin(p \theta - \frac{2\pi}{3}) & \sin(p \theta + \frac{2\pi}{3}) \end{bmatrix}
\end{equation}
is the Park-Clarke transformation.

Now, assume that the power converter is a voltage source inverter in order to set the stator voltage. Suppose vector $\bm v =[v_d \;\; v_q]^T$ represents the stationary dq-axes phase voltages. Then, by virtue of the dynamics model of motor torque \eqref{eq:dottau}, we propose the following linearization control law in terms of control inputs $u$ and $\bm z$
\begin{equation}\label{eq:v_lin}
\bm v = \frac{\bm b}{\| \bm b \|^2} (u - \phi) + \bm z,
\end{equation}
where scaler $u$ can take any value but $\bm z$ is specifically defined to satisfy the following constraint
\begin{equation} \label{eq:bTz=0}
\bm b^T \bm z =0.
\end{equation}
One can readily verify that  substitution of the voltage control law \eqref{eq:v_lin} into the time-derivative of motor torque in \eqref{eq:dottau} yields the following first-order linear system
\begin{align} \notag
\tau + \mu \dot{\tau} &= \bm b^T \frac{\bm b}{\| \bm b \|^2} (u - \phi) + \bm b^T\bm z + \phi \\\label{eq:dtau=u}
&= u
\end{align}
It is apparent from \eqref{eq:dtau=u} that input $\bm z$ does affect the motor torque generation and hence control input $u$ is exclusively responsible for the torque production. Consequently, we treat $u$ and $\bm z$ as the torque control input and energy minimizer control input, respectively. Equation \eqref{eq:v_lin} can be interpreted  as a transformation from the auxiliary inputs $u$ and $\bm z$ to the dq-axes phase voltages. Control input $u$ constitutes the reference torque, i.e.,  $\tau + \mu \dot{\tau}=u$. However, as  will be shown in the following section,  the other control input $\bm z$ can be utilized to minimize power dissipation due to the copper loss.

\section{Optimal control to minimize energy consumption} \label{sec:minimize_energy}
\subsection{Optimal control}  \label{sec:opt_cntr}
By substituting the linearization control law \eqref{eq:v_lin} into the machine voltage equations \eqref{eq:vdq}, we arrive at the following time-varying linear system describing the  dynamics of the currents in response to the control inputs $u$ and $\bm z$
\begin{align} \label{eq:diffi_2inputs}
\frac{d \bm i}{dt} &= \bm L^{-1} \Big( \frac{\bm b}{\| \bm b \|^2} (u(t) - \phi) + \bm h+ \bm z \Big), \\
& = \bm f (\bm i, t, \bm z)
\end{align}
where $\bm L=\mbox{diag}\{L_d, \; L_q \}$ and vector $\bm h$ is defined as
\begin{equation} \notag
\bm h(\bm i, \omega) = \begin{bmatrix} - R i_d + L_q i_q \omega \\ - R i_q + L_d i_d \omega - \psi \omega  \end{bmatrix}
\end{equation}

The cost function to minimize is  power dissipation $J=\| \bm i \|^2$ over interval $h$, i.e.,
\begin{equation} \label{eq:cost}
\int_t^T \| \bm i(\zeta) \|^2 d \zeta
\end{equation}
where $T= t + h$ is the terminal time of the system. Then, the {\em Hamiltonian} function can be constructed from \eqref{eq:diffi_2inputs} and \eqref{eq:cost} as
\begin{align}\notag
H &= J + \bm\lambda^T \bm f \\ \notag
&= \| \bm i \|^2 + \bm\lambda^T \frac{d \bm i}{dt} \\ \label{eq:H1}
&= \| \bm i \|^2 + \bm\lambda^T  \bm L^{-1} \Big( \frac{\bm b}{\| \bm b \|^2} (u - \phi) +\bm h + \bm z  \Big),
\end{align}
where $\bm\lambda \in \mathbb{R}^2$ is the costate vector. The  optimality condition stipulates that the time-derivative of costate satisfies
\begin{equation} \label{eq:costates_derivative}
\dot{\bm\lambda} = - \frac{\partial H}{\partial \bm i},
\end{equation}
and the transversally condition dictates
\begin{equation}\label{eq:lamT}
\bm\lambda(T) =0
\end{equation}
Therefore, using using the Hamiltonian expression from \eqref{eq:H1} in \eqref{eq:costates_derivative}, one can show that evolution of the system costate is governed by the following time-varying differential equation
\begin{equation} \label{eq:difp}
\frac{d}{dt} \bm\lambda =  \bm A^T \bm\lambda - 2 \bm i
\end{equation}
where
\begin{align*}
\bm A &= \bm L^{-1} \Big(-(u - \phi) \frac{\partial}{\partial \bm i} \frac{\bm b}{\| \bm b \|^2} + \frac{\bm b}{\| \bm b \|^2}\frac{\partial \phi^T}{\partial \bm i} - \frac{\partial \bm h}{\partial \bm i} \Big)\\
&= (u-\phi)\bm\Lambda + \bm\Gamma
 \end{align*}
and
\begin{align*}
\bm\Lambda &= \frac{3 p \eta L_d}{2R \| \bm b \|^4} \bm L^{-1} \begin{bmatrix} 2 b_d b_q & b_q^2 - b_d^2 \\ b_d^2 - b_q^2 & 2 b_d b_q \end{bmatrix} \\
\bm\Gamma & = \bm L^{-1} \Big( \frac{\bm b}{\| \bm b \|^2} \frac{\partial \phi^T}{\partial \bm i} - \frac{\partial \bm h}{\partial \bm i} \Big) \\
\frac{\partial \phi}{\partial \bm i} &= \frac{3p}{2}\begin{bmatrix} \mu \omega \psi  + \eta L_q i_q  - 2 \frac{\omega}{R} \eta L_d^2 i_d \\ - 2\omega \mu  \eta L_q i_q + \eta L_q i_d  \end{bmatrix}\\
\frac{\partial \bm h}{\partial \bm i} &= \begin{bmatrix}-R & L_q \omega \\ L_d \omega & -R\end{bmatrix}
\end{align*}

Dynamics equation \eqref{eq:difp}  can be used as an observer to estimate the costate $\bm\lambda$.  We can write the equivalent discrete-time model of the continuous system \eqref{eq:difp} as
\begin{equation}
\frac{1}{h}(\bm\lambda_{k+1}  - \bm\lambda_k) = \bm A_k^T \bm\lambda_k - 2 \bm i_k
\end{equation}
The boundary condition can be inferred from \eqref{eq:lamT} as $\bm\lambda_{k+1}=0$. Using the latter identity in the above equation, we get
\begin{equation} \label{eq:lambk}
\bm\lambda_k = 2(\frac{1}{h} \bm I + \bm A_k^T)^{-1} \bm i_k
\end{equation}
Notice that computation of the costate from   \eqref{eq:lambk} does not involve its time-history. Therefore, for the sake of notational simplicity, we will drop the $_k$ subscript of the variables in the following analysis without causing ambiguity.

Moreover, according to the {\em Pontryagin's minimum principle}, the optimal control input minimizes the Hamiltonian over  the set of all permissible controls and over optimal trajectories of the state $\bm i^*$ and costate $\bm\lambda^*$, i.e.,
\begin{equation} \label{eq:minH}
\bm z = \mbox{arg} \min_{\bm z \in \mathcal{Z}} H(\bm i^*, \bm\lambda^*, \bm z)
\end{equation}
where
\begin{equation} \label{eq:Zclass}
\mathcal{Z}=\{\bm z : \;   \bm z \perp \bm b \}
\end{equation}
is the set of admissible control inputs. Recall that the constraint in \eqref{eq:Zclass} ensures that the energy minimizer control input $\bm z$ does not contribute to the motor torque generation. It can be inferred from the expression of Hamiltonian \eqref{eq:H1} and identity   \eqref{eq:bTz=0} that \eqref{eq:minH} is tantamount to minimizing $\bm\lambda^T \bm L^{-1}  \bm z$  subject to the equality and inequality constraints of admissible $ \bm b^T \bm z =0$. Therefore, the problem of finding optimal permissible $\bm z$  can be {\em transcribed}
to the following {\em constrained linear programming}
\begin{align}\label{eq:LP}
\mbox{minimum} \;\;\;   &\bm\lambda^T \bm L^{-1} \bm z\\ \notag
\mbox{subject to}  \quad  & \bm b^T  \bm z=0
\end{align}
Let us define  projection matrix
\begin{equation} \label{eq:B}
\bm B = \bm I - \frac{\bm b \bm b^T}{\|\bm b \|^2},
\end{equation}
which projects vector from $\mathbb{R}^2$ to a vector space perpendicular to $\bm b$, i.e., $\bm B \bm b = \bm 0$. Then, it can be inferred from \eqref{eq:LP} and \eqref{eq:B} that optimal $\bm z$ should be aligned with vector $\bm B \bm L^{-1} \bm\lambda$ in opposite direction, i.e.,
\begin{equation} \label{eq:zopt}
\bm z = - \gamma \bm B \bm L^{-1} \bm\lambda
\end{equation}
and $\gamma>0$ is a positive scaler. Finally, the energy minimizer control command is obtained from \eqref{eq:zopt} and \eqref{eq:lambk}.

\subsection{Voltage saturation limit}   \label{sec:voltage_limit}
A power converter has a maximum voltage limit, which is the bus voltage.
The variable $\gamma$ in \eqref{eq:zopt} should take  a large value  in order to minimize energy consumption. However, a very large $\gamma$ tends to increase the inverter voltage towards the saturation limit. Therefore, the value of $\gamma$ should be selected large as much as possible as long as the voltage vector does not reach its saturation limit, i.e.,
\begin{equation} \label{eq:v_max}
\| \bm v \| \leq v_{\rm max}
\end{equation}
where $v_{\rm max}$ is the maximum phase voltage magnitude \cite{Sue-Pan-2008}. From \eqref{eq:v_lin}, we can say
\begin{equation} \label{eq:v_norm}
\| \bm v \|^2 = \frac{( u - \phi )^2}{\| \bm b \|^2}  + \| \bm z \|^2
\end{equation}
In view of \eqref{eq:v_norm} and \eqref{eq:v_max}, the maximum allowable magnitude of control input $\bm z$ is
\begin{equation} \label{eq:z_abs}
\| \bm z \| \leq  z_{\rm max}
\end{equation}
where the limits of the auxiliary control input $\bm z$ is
\begin{equation} \label{eq:zsqrt}
\quad z_{\rm max} = \sqrt{ v_{\rm max}^2  - (u - \phi)^2/ \| \bm b \|^2}
\end{equation}
In other words, the set of admissible control input \eqref{eq:Zclass} in the presence of voltage saturation limit becomes
\begin{equation} \label{eq:Zclass2}
\mathcal{Z}=\{\bm z : \;   \bm z \perp \bm b \;\; \wedge \;\; \| \bm z\| \leq z_{\rm max} \}
\end{equation}

By virtue of \eqref{eq:zopt}, \eqref{eq:zsqrt}, and \eqref{eq:z_abs}, we can conclude the optimal value of variable $\gamma$ minimizing power losses given the voltage limit $v_{\rm max}$ to be
\begin{equation} \label{eq:gamma}
\gamma = \frac{\sqrt{ v_{\rm max}^2  - (u - \phi)^2/ \| \bm b \|^2}}{\| \bm B \bm L^{-1} \bm\lambda \|}
\end{equation}

\subsection{Torque limit} \label{sec:torque_limit}
It should be pointed out that the expression under the square-root in  \eqref{eq:zsqrt} must be positive  to ensure  real-valued  solution for the control input $\bm z$. That requires
\begin{equation} \notag
v_{\rm max}\| \bm b \| \geq | u - \phi |.
\end{equation}
Therefore, the value of  the torque command should be within the following bands
\begin{equation}
u_{\rm min} \leq u \leq u_{\rm max},
\end{equation}
where $u_{\rm min} = \phi - \| \bm b \| v_{\rm max}$ and $u_{\rm max} = \phi + \| \bm b \| v_{\rm max}$
are the lower- and upper-bounds. Therefore, the torque control input $u$ can be modified according to the following to ensure feasible solution
\begin{equation} \label{eq:ulimits}
u = \left\{ \begin{array}{ll} u_{\rm max} \quad & \mbox{if} \quad u > u_{\rm max} \\
u \quad & \mbox{if} \quad   u_{\rm min} \leq u \leq u_{\rm max} \\
u_{\rm min} \quad & \mbox{if} \quad  u < u_{\rm min} \end{array} \right.
\end{equation}

Now with $u$ and $\bm z$ in hand, one may use \eqref{eq:v_lin} to calculate stationary dq-axes phase voltages. Finally, the inverter phase voltages can be obtained from the inverse Park-Clarke transform, i.e.,
\begin{equation}\label{eq:invPark}
\begin{bmatrix} v_a \\ v_b \\ v_c \end{bmatrix} = \bm K^{-1}(\theta) \begin{bmatrix} v_d \\ v_q \end{bmatrix}
\end{equation}
where
\begin{equation} \label{eq:Kinv}
\bm K^{-1}(\theta) = \begin{bmatrix} \cos(p \theta) &  \sin(p \theta)\\  \cos(p \theta - \frac{2\pi}{3}) & \sin(p \theta- \frac{2\pi}{3}) \\  \cos(p \theta + \frac{2\pi}{3}) &  \sin(p \theta + \frac{2\pi}{3}) \end{bmatrix}.
\end{equation}

\subsection{Torque bandwidth} \label{sec:Bandwidth}
The linearization control voltage vector $\bm v$ is computed based on auxiliary input $u(t)$ and the optimizing control $\bm z$. Upon applying the optimal feedback linearization control to the motor, the input/output of the closed-loop system in the Laplace domain is simply given by
\begin{equation} \label{eq:tau/u}
\frac{\tau(s)}{u(s)} = \frac{1}{\mu s + 1}
\end{equation}
where $s$ is the Laplace variable and  $\mu$ is the machine time-constant as previously defined. Since the linearized system \eqref{eq:tau/u} is strictly stable, the proposed feedback linearization control scheme is inherently robust without recurring to external torque feedback loop. Nevertheless in order to increase the bandwidth  of the linearized system, one may consider a PI feedback loop  closed around the linearized system,  where $\tau^*$ is the desired input torque and $\tau(\bm i)$ is the motor torque estimated from d-q currents according to \eqref{eq:tau}.

In summary, the control algorithm may proceed with the following steps:
\begin{enumerate}
\item Acquire data pertaining to shaft position and speed, and the phase currents from sensors. Then, compute dq currents from Park-Clarke transform \eqref{eq:Park}.
\item Estimate motor torque according to \eqref{eq:tau} and then obtain $u$ from the PI feedback loop.
\item Given torque control input $u$ and maximum voltage limit $v_{\rm max}$, limit the magnitude of the command according to \eqref{eq:ulimits}.
\item Use \eqref{eq:lambk} to estimate the costate vector $\bm\lambda$.
\item Compute the energy minimizer control input $\bm z$ from \eqref{eq:zopt} and \eqref{eq:gamma}.
\item With $u$ and $\bm z$ in hand, use the linearization transformation \eqref{eq:v_lin} to obtain the stationary dq-axes phase voltages. Then, compute the inverter phase voltages from the inverse Park-Clarke transform \eqref{eq:Kinv}.
\end{enumerate}

\section{Conclusions}
The Hamiltonian of optimal control theory in conjunction with the Pontryagin's minimum principle have been rigorously applied to derive  an optimal feedback-linearization control scheme  for energy-efficient and accurate torque control of PMSMs subject to time-varying operational speed/torque and voltage limit. Analytical solution for the optimal control problem has been found based on the {\em maximum principle} formulation for convenience of real-time implementation. The important feature of the optimal controller was that it could admit non-constant reference torques (or velocity) as oppose to the conventional regulation control approaches requiring the motor to operate at constant torque or velocity (regulation). The optimal controller could achieve accurate torque tracking with minimize power consumption given inverter voltage limit.

\bibliographystyle{IEEEtran}

\begin{thebibliography}{10}

\bibitem{Straete-Degezelle-DeSchutter-1998}
H.~van~de Straete, P.~Degezelle, J.~De~Schutter, and R.~J.~M. Belmans, ``Servo
  motor selection criterion for mechatronic applications,'' \emph{Mechatronics,
  IEEE/ASME Transactions on}, vol.~3, no.~1, pp. 43--50, Mar 1998.

\bibitem{Aghili-Buehler-Hollerbach-2007}
F.~Aghili, J.~M.~Hollerbach, and M.~Buehler, ``A Modular and High-Precision Motion Control System with an Integrated Motor,'' {\em {IEEE/ASME} Trans. on Mechatronics\/}, Vol.~12, No.~3, June 2007, pp.~317--329.

\bibitem{Zhong-Rahman-Hu-1997}
L.~Zhong, M.~Rahman, W.~Y. Hu, and K.~W. Lim, ``Analysis of direct torque
  control in permanent magnet synchronous motor drives,'' \emph{Power
  Electronics, IEEE Transactions on}, vol.~12, no.~3, pp. 528--536, May 1997.

\bibitem{Buja-Kazmierkowski-2004}
G.~S. Buja and M.~P. Kazmierkowski, ``Direct torque control of pwm inverter-fed
  ac motors - a survey,'' \emph{IEEE Transactions on Industrial Electronics},
  vol.~51, no.~4, pp. 744--757, Aug 2004.

\bibitem{Aghili-Buehler-Hollerbach-2001}
F.~Aghili, J.~M.~Hollerbach, and M.~Buehler, ``Motion Control Systems
  with {$H_{\infty}$} Positive Joint Torque Feedback,'' {\em IEEE Trans. Control
  Systems Technology\/}, Vol.~9, No.~5, 2001, pp.~685--695.


\bibitem{Grenier-Dessaint-Akhrif-1997}
D.~Grenier, L.-A. Dessaint, O.~Akhrif, Y.~Bonnassieux, and B.~Le~Pioufle,
  ``Experimental nonlinear torque control of a permanent-magnet synchronous
  motor using saliency,'' \emph{Industrial Electronics, IEEE Transactions on},
  vol.~44, no.~5, pp. 680--687, Oct 1997.

\bibitem{Aghili-2015a}
F.~Aghili, ``Energy-efficient and fault-tolerant control of multiphase
  nonsinusoidal pm synchronous machines,'' \emph{Mechatronics, IEEE/ASME
  Transactions on}, vol.~20, no.~6, pp. 2736--2751, Dec 2015.

\bibitem{Cheng-Tesch-2010}
B.~Cheng and T.~R. Tesch, ``Torque feedforward control technique for
  permanent-magnet synchronous motors,'' \emph{IEEE Transactions on Industrial
  Electronics}, vol.~57, no.~3, pp. 969--974, March 2010.

\bibitem{Aghili-2017b}
F.~Aghili, ``Optimal Feedback Linearization Control of Interior PM
  Synchronous Motors Subject to Time-Varying Operation Conditions Minimizing
  Power Loss,'' {\em IEEE Transactions on Industrial Electronics\/}, Vol.~65,
  No.~7, July 2018, pp.~5414--5421.

\bibitem{Sue-Pan-2008}
S.~M. Sue and C.~T. Pan, ``Voltage-constraint-tracking-based field-weakening
  control of ipm synchronous motor drives,'' \emph{IEEE Transactions on
  Industrial Electronics}, vol.~55, no.~1, pp. 340--347, Jan 2008.

\bibitem{Bodson-Chiasson-Novotnak-1993}
M.~Bodson, J.~Chiasson, R.~Novotnak, and R.~Rekowski, ``High-performance
  nonlinear feedback control of a permanent magnet stepper motor,''
  \emph{Control Systems Technology, IEEE Transactions on}, vol.~1, no.~1, pp.
  5--14, Mar 1993.

\bibitem{Kaddouri-Akhrif-1994}
A.~Kaddouri, O.~Akhrif, H.~Le-Huy, and M.~Ghribi, ``Nonlinear feedback control
  of a permanent magnet synchronous motors,'' in \emph{Electrical and Computer
  Engineering, 1994. Conference Proceedings. 1994 Canadian Conference on}, Sep
  1994, pp. 77--80 vol.1.

\bibitem{Takahashi-Noguchi-1986}
I.~Takahashi and T.~Noguchi, ``A new quick-response and high-efficiency control
  strategy of an induction motor,'' \emph{Industry Applications, IEEE
  Transactions on}, vol. IA-22, no.~5, pp. 820--827, Sept 1986.

\bibitem{Zhang-Zhu-2011}
Y.~Zhang and J.~Zhu, ``Direct torque control of permanent magnet synchronous
  motor with reduced torque ripple and commutation frequency,'' \emph{Power
  Electronics, IEEE Transactions on}, vol.~26, no.~1, pp. 235--248, Jan 2011.

\bibitem{Gulez-Adam-Pastaci-2007}
K.~Gulez, A.~Adam, and H.~Pastaci, ``A novel direct torque control algorithm
  for ipmsm with minimum harmonics and torque ripples,'' \emph{Mechatronics,
  IEEE/ASME Transactions on}, vol.~12, no.~2, pp. 223--227, April 2007.

\bibitem{Cho-Yongsoo-Kim-2013}
Y.~Cho, D.-H. Kim, K.-B. Lee, Y.~I. Lee, and J.-H. Song, ``Torque ripple
  reduction and fast torque response strategy of direct torque control for
  permanent-magnet synchronous motor,'' in \emph{Industrial Electronics (ISIE),
  2013 IEEE International Symposium on}, May 2013, pp. 1--6.

\bibitem{Chen-Chin-2000}
J.-J. Chen and K.-P. Chin, ``Automatic flux-weakening control of permanent
  magnet synchronous motors using a reduced-order controller,'' \emph{Power
  Electronics, IEEE Transactions on}, vol.~15, no.~5, pp. 881--890, Sep 2000.

\bibitem{Zhu-Chen-How-2000}
Z.~Zhu, Y.~Chen, and D.~Howe, ``Online optimal flux-weakening control of
  permanent-magnet brushless ac drives,'' \emph{Industry Applications, IEEE
  Transactions on}, vol.~36, no.~6, pp. 1661--1668, Nov 2000.

\bibitem{Chen-Chin-2003}
J.-J. Chen and K.-P. Chin, ``Minimum copper loss flux-weakening control of
  surface mounted permanent magnet synchronous motors,'' \emph{Power
  Electronics, IEEE Transactions on}, vol.~18, no.~4, pp. 929--936, July 2003.

\bibitem{Chiang-Cheng-Hsum-2015}
H.-H. Chiang, K.-C. Hsu, and I.-H. Li, ``Optimized adaptive motion control
  through an sopc implementation for linear induction motor drives,''
  \emph{Mechatronics, IEEE/ASME Transactions on}, vol.~20, no.~1, pp. 348--360,
  Feb 2015.

\bibitem{Chang-Low-1994}
K.-T. Chang, T.-S. Low, and T.-H. Lee, ``An optimal speed controller for
  permanent-magnet synchronous motor drives,'' \emph{IEEE Transactions on
  Industrial Electronics}, vol.~41, no.~5, pp. 503--510, Oct 1994.

\bibitem{Do-Kwak-2014}
T.~D. Do, S.~Kwak, H.~H. Choi, and J.~W. Jung, ``Suboptimal control scheme
  design for interior permanent-magnet synchronous motors: An sdre-based
  approach,'' \emph{IEEE Transactions on Power Electronics}, vol.~29, no.~6,
  pp. 3020--3031, June 2014.

\bibitem{El-Sousy-2013}
F.~F.~M. El-Sousy, ``Intelligent optimal recurrent wavelet elman neural network
  control system for permanent-magnet synchronous motor servo drive,''
  \emph{IEEE Transactions on Industrial Informatics}, vol.~9, no.~4, pp.
  1986--2003, Nov 2013.

\bibitem{Cheema-Fletcher-2016}
M.~A.~M. Cheema, J.~E. Fletcher, D.~Xiao, and M.~F. Rahman, ``A linear
  quadratic regulator-based optimal direct thrust force control of linear
  permanent-magnet synchronous motor,'' \emph{IEEE Transactions on Industrial
  Electronics}, vol.~63, no.~5, pp. 2722--2733, May 2016.

\bibitem{Guerrero-Leetmaa-Briz-2003}
J.~M. Guerrero, M.~Leetmaa, F.~Briz, A.~Zamarron, and R.~D. Lorenz, ``Inverter
  nonlinearity effects in high frequency signal injection-based, sensorless
  control methods,'' in \emph{38th IAS Annual Meeting on Conference Record of
  the Industry Applications Conference, 2003.}, vol.~2, Oct 2003, pp.
  1157--1164 vol.2.

\bibitem{Kazmierkowski-Malesani-1998}
M.~P. Kazmierkowski and L.~Malesani, ``Current control techniques for
  three-phase voltage-source pwm converters: a survey,'' \emph{IEEE
  Transactions on Industrial Electronics}, vol.~45, no.~5, pp. 691--703, Oct
  1998.

\bibitem{Sepe-Lang-1994}
R.~B. Sepe and J.~H. Lang, ``Inverter nonlinearities and discrete-time vector
  current control,'' \emph{IEEE Transactions on Industry Applications},
  vol.~30, no.~1, pp. 62--70, Jan 1994.

\bibitem{Aghili-2011m}
F.~Aghili, ``Optimal and fault-tolerant torque control of servo motors subject
  to voltage and current limits,'' \emph{{IEEE} Trans. on Control System
  Technology}, vol.~21, no.~4, pp. 1440--1448, July 2013.

\bibitem{Krneta-Antic-Stojanovic-2005}
R.~Krneta, S.~Antic, and D.~Stojanovic, ``Recursive least squares method in
  parameters indentification of dc motors models,'' \emph{Facta Universitatis
  Series : Electronics and Energetics}, vol.~18, no.~3, 2005.

\bibitem{Aghili-2008a}
F.~Aghili, ``Adaptive reshaping of excitation currents for accurate torque
  control of brushless motors,'' \emph{IEEE Trans. on Control System
  Technologies}, vol.~16, no.~2, pp. 356--364, Mar. 2008.

\bibitem{Obeidat-Wang-Lin-2013}
M.~A. Obeidat, L.~Y. Wang, and F.~Lin, ``Real-time parameter estimation of pmdc
  motors using quantized sensors,'' \emph{IEEE Transactions on Vehicular
  Technology}, vol.~62, no.~7, pp. 2977--2986, Sept 2013.

\bibitem{Li-Elbuluk-2001}
C.~Li and M.~Elbuluk, ``A sliding mode observer for sensorless control of
  permanent magnet synchronous motors,'' in \emph{Conference Record of the 2001
  IEEE Industry Applications Conference. 36th IAS Annual Meeting (Cat.
  No.01CH37248)}, vol.~2, Sept 2001, pp. 1273--1278 vol.2.

\bibitem{Reichhartinger-2010}
M.~Reichhartinger and M.~Horn, ``Sliding-mode control of a permanent-magnet
  synchronous motor with uncertainty estimation,'' in \emph{International
  Journal of Electrical and Computer Engineering}, vol.~4, no.~11, 2010, pp.
  1637--1640.

\bibitem{Qiao-Shi-Wang-2013}
Z.~Qiao, T.~Shi, Y.~Wang, Y.~Yan, C.~Xia, and X.~He, ``New sliding-mode
  observer for position sensorless control of permanent-magnet synchronous
  motor,'' \emph{IEEE Transactions on Industrial Electronics}, vol.~60, no.~2,
  pp. 710--719, Feb 2013.

\bibitem{Jovanovic-Betz-1999}
M.~G. Jovanovic and R.~E. Betz, ``Optimal torque controller for synchronous
  reluctance motors,'' \emph{IEEE Transactions on Energy Conversion}, vol.~14,
  no.~4, pp. 1088--1093, Dec 1999.

\bibitem{Aghili-Buehler-Hollerbach-2000b}
F.~Aghili, M.~Buehler, and J.~M. Hollerbach, ``Optimal commutation laws in the
  frequency domain for {PM} synchronous direct-drive motors,'' \emph{IEEE
  Transactions on Power Electronics}, vol.~15, no.~6, pp. 1056--1064, Nov.
  2000.

\bibitem{Mademlis-Xypteras-2000}
C.~Mademlis, J.~Xypteras, and N.~Margaris, ``Loss minimization in surface
  permanent-magnet synchronous motor drives,'' \emph{IEEE Transactions on
  Industrial Electronics}, vol.~47, no.~1, pp. 115--122, Feb 2000.

\bibitem{Aghili-Buehler-Hollerbach-2003}
F.~Aghili, M.~Buehler, and J.~M. Hollerbach, ``Experimental characterization
  and quadratic programming-based control of brushless-motors,'' \emph{IEEE
  Trans. on Control Systems Technology}, vol.~11, no.~1, pp. 139--146, 2003.

\bibitem{Kestelyn-Semail-2011}
X.~Kestelyn and E.~Semail, ``A vectorial approach for generation of optimal
  current references for multiphase permanent-magnet synchronous machines in
  real time,'' \emph{Industrial Electronics, IEEE Transactions on}, vol.~58,
  no.~11, pp. 5057--5065, Nov 2011.

\bibitem{Jahns-Kliman-1986}
T.~M. Jahns, G.~B. Kliman, and T.~W. Neumann, ``Interior permanent-magnet
  synchronous motors for adjustable-speed drives,'' \emph{IEEE Transactions on
  Industry Applications}, vol. IA-22, no.~4, pp. 738--747, July 1986.

\bibitem{Nakai-Ohtani-Satoh-2001}
H.~Nakai, H.~Ohtani, E.~Satoh, and Y.~Inaguma, ``Development and testing of the
  torque control for the permanent-magnet synchronous motor,'' in
  \emph{Industrial Electronics Society, 2001. IECON '01. The 27th Annual
  Conference of the IEEE}, vol.~2, 2001, pp. 1463--1468 vol.2.

\bibitem{Morimoto-Hatanaka-Tong-1991}
S.~Morimoto, K.~Hatanaka, Y.~Tong, Y.~Takeda, and T.~Hirasa, ``High performance
  servo drive system of salient pole permanent magnet synchronous motor,'' in
  \emph{Conference Record of the 1991 IEEE Industry Applications Society Annual
  Meeting}, Sept 1991, pp. 463--468 vol.1.

\bibitem{Morimoto-Sanada-Takeda-1994}
S.~Morimoto, M.~Sanada, and Y.~Takeda, ``Wide-speed operation of interior
  permanent magnet synchronous motors with high-performance current
  regulator,'' \emph{IEEE Transactions on Industry Applications}, vol.~30,
  no.~4, pp. 920--926, Jul 1994.

\bibitem{Nakai-Ohtani-2005}
H.~Nakai, H.~Ohtani, E.~Satoh, and Y.~Inaguma, ``Development and testing of the
  torque control for the permanent-magnet synchronous motor,'' \emph{IEEE
  Transactions on Industrial Electronics}, vol.~52, no.~3, pp. 800--806, June
  2005.

\bibitem{Rang-Lim-Nam-2004}
G.~Rang, J.~Lim, K.~Nam, H.-B. Ihm, and H.-G. Kim, ``A mtpa control scheme for
  an ipm synchronous motor considering magnet flux variation caused by
  temperature,'' in \emph{Applied Power Electronics Conference and Exposition,
  2004. APEC '04. Nineteenth Annual IEEE}, vol.~3, 2004, pp. 1617--1621 Vol.3.

\bibitem{Sebastian-Slemon-Rahman-1987}
T.~Sebastian, G.~Slemon, and M.~Rahman, ``Modelling of permanent magnet
  synchronous motors,'' \emph{IEEE Transactions on Magnetics}, vol.~22, no.~5,
  pp. 1069--1071, Sep 1986.

\bibitem{Pillay-Krishnan-1989}
P.~Pillay and R.~Krishnan, ``Modeling, simulation, and analysis of
  permanent-magnet motor drives. i. the permanent-magnet synchronous motor
  drive,'' \emph{IEEE Transactions on Industry Applications}, vol.~25, no.~2,
  pp. 265--273, Mar 1989.

\bibitem{Gebregergis-Chowdhury-2015}
A.~Gebregergis, M.~H. Chowdhury, M.~S. Islam, and T.~Sebastian, ``Modeling of
  permanent-magnet synchronous machine including torque ripple effects,''
  \emph{IEEE Transactions on Industry Applications}, vol.~51, no.~1, pp.
  232--239, Jan 2015.

\bibitem{Aghili-2006}
F.~Aghili, ``A mechatronic testbed for revolute-joint prototypes of a
  manipulator,'' \emph{IEEE Trans. on Robotics}, vol.~22, no.~6, pp.
  1265--1273, Dec. 2006.

\end{thebibliography}

\end{document}